\documentstyle[12pt,epsf]{article}
\if@twoside
 \else
 \fi
 \parskip \medskipamount
\begin{document}

\hfill    NBI-HE-97-18

\hfill May 1997

\begin{center}
\vspace{24pt}
{\large \bf Improved determination of the classical\\
        sphaleron transition rate}

\vspace{36pt}

{\sl J. Ambj\o rn}\footnote{E-mail: ambjorn@nbi.dk}
 and {\sl A. Krasnitz}\footnote{E-mail: krasnitz@nbi.dk}  \\

\vspace{24pt}
 The Niels Bohr Institute,\\
Blegdamsvej 17, \\
DK-2100 Copenhagen \O , Denmark.\\

\end{center}
\vspace{24pt}

\vfill

\begin{center}
{\bf Abstract}
\end{center}
\vspace{12pt}

\noindent
We determine the sphaleron transition rate using real time 
lattice simulations of the classical system. An improved 
definition of the lattice topological charge allows us
to obtain a more reliable estimate of the transition rate.
For an SU(2) Yang-Mills-Higgs system in the broken phase we find
the transition rate to be strongly suppressed, and we have observed no
sphaleron transitions in the range of coupling constants used. For a 
pure SU(2) Yang-Mills system in large volumes the rate 
behaves as $\kappa (\alpha_{\rm w}T)^4$, with $\kappa$ slightly
decreasing as the lattice spacing is reduced.
If the lattice size is reduced to about twice the magnetic screening length,
the rate is suppressed by finite-size effects, and $\kappa$ is approximately
proportional to the lattice spacing.
Our rate measurements are supplemented
by analysis of gauge field correlation functions in the Coulomb gauge.

\vspace{12pt}

\bigskip



\vfill

\newpage

\section{Introduction}
Knowledge of the high-temperature baryon-number violation rate in the 
electroweak theory may be a key to origins of baryon matter. 
For temperatures close 
to and above the electroweak phase transition this
essentially nonperturbative quantity can only be determined under two 
simplifying assumptions that {\it (a)} fermion degrees of freedom can be 
taken into account through modification of coupling constants,
and that {\it (b)} the classical approximation is valid. The rate is
then found by numerically following the real-time evolution of
the Chern-Simons number in the classical (lattice) cutoff theory.
This line of research was initiated in \cite{janetal}, and  
the last two years have seen much numerical effort to
improve the simulations, 
as well as theoretical examination of the simplifying assumptions
(mostly the second one). 
In our large-scale numerical simulation of the SU(2) Yang-Mills theory 
\cite{su2rate} we found that, with the standard approximate discretization of
topological charge density, the classical Chern-Simons number diffusion 
rate $\Gamma$ is finite in the continuum limit:
\begin{equation} \Gamma=\kappa\left(\alpha_{\rm w}T\right)^4. 
\label{t4law}\end{equation}
We also estimated the value of the dimensionless coefficient $\kappa$ in front
of the $T^4$ law (\ref{t4law}): $\kappa=1.09\pm 0.05$. Moore \cite{mooremu}
employed a different method for numerical determination of $\Gamma$, and his
results agreed with ours. Tang and Smit \cite{ts} measured $\Gamma$ in the
SU(2) Yang-Mills-Higgs theory. In the low-temperature Higgs phase of the theory
$\Gamma$ is theoretically expected to be suppressed by the Boltzmann factor
of the sphaleron potential barrier \cite{armcl}, rendering Chern-Simons number
diffusion inobservable in a realistic numerical simulation. Surprisingly, no
such strong suppression was found in \cite{ts}. Even more dramatic deviation
from the sphaleron regime was found by Moore and Turok, raising suspicion that
the residual diffusion in the Higgs phase is an artifact of the classical
lattice theory \cite{mt}. Arnold, Son and Yaffe \cite{asy} (ASY in the 
following), and later Arnold \cite{arnonly} further questioned the validity
of the classical result for $\Gamma$, arguing that the classical $\Gamma$ must
be proportional to the lattice spacing and therefore vanish in the continuum 
limit. More recently, Moore and Turok \cite{mt2}, employing an improved 
definition of Chern-Simons number, found that the transition rate in the
lower-temperature phase was zero within the measurement error, as expected 
from the sphaleron approximation. They also detected a dependence of the rate
on the lattice spacing, but that dependence was considerably milder than the
prediction \cite{asy,arnonly} and left ample room for the existence of
continuum limit.

Our present work is an effort to further examine the classical result for 
$\Gamma$ and to identify its dependence  on the lattice discretization.
Our new results show that, with the standard lattice
definition of topological charge density, there are spurious contributions
to $\Gamma$, directly related to the well-known difficulty with defining
topological quantities in a compact lattice gauge theory. We 
introduce a new definition of the lattice topological charge, designed to 
strongly suppress these artifacts. With most of the artifacts removed, $\Gamma$
is indeed inobservably low in the Higgs phase of the Yang-Mills-Higgs theory.
In the pure Yang-Mills theory, the use of the standard definition of topological
charge leads to a systematic error in $\Gamma$. Arnold, Son, and Yaffe
conjectured that this systematic error creates an illusion of a 
$\kappa$ independent of the lattice spacing, while the correct lattice rate
vanishes in the continuum limit. We do find a tendency of the rate to decrease
with the lattice spacing, but this tendency does not appear to be strong enough
to rule out the existence of the continuum limit. There is nevertheless one
instance in which the rate exhibits proportionality to the lattice spacing in a
wide range of the latter: the Yang-Mills theory in a small volume, where
finite-size effects give rise to a sphaleron barrier. In that case, however,
there remains a possibility that the continuum rate exists, but is very small
due to an exponential suppression of barrier configurations.

The plan of this paper is as follows. We begin by explaining the cooling method
for measuring lattice topological charge in Section \ref{topch}. This method 
is used to place an upper bound on the rate in the low-temperature phase of the
Yang-Mills-Higgs system (Section \ref{ymh}) and to estimate the rate in the 
large-volume limit of the Yang-Mills theory (Section \ref{ymbig}). We also 
consider sphaleron transitions in the Yang-Mills theory in a small volume
(Section \ref{ymtiny}). A closer look at the ASY scenario is taken in Section
\ref{glue}, where we study the dynamics of gauge fields in the Coulomb gauge.
A discussion of the emerging picture is given in Section \ref{disc}.

\section{Cooling method for the lattice topological charge}\label{topch}
Consider the standard lattice approximation for the topological charge per unit
time
\begin{equation}
\dot N_{\rm CS}={i\over{32\pi^2}}\sum_{j,n}\left(\overline{E}^\alpha_{j,n}
+E^\alpha_{j-n,n}\right)
\sum_{\Box_{j,n}}{\rm Tr}\left(U_{\Box_{j,n}}\sigma^\alpha\right),
\label{standard}\end{equation}
where
$$\overline{E}^\alpha_{j,n}\equiv {1\over 2}E^\beta_{j,n}{\rm Tr}
\left(\sigma^\alpha U_{j,n}\sigma^\beta U_{j,n}^\dagger\right).$$
Using the stndard notation, we assign to every lattice site $j$
link matrices $U_{j,n}$ in the fundamental representation of SU(2) and lattice 
analogs of color electric fields
$E_{j,n}$, three of each, corresponding to the three positive lattice
directions $n$. The usual plaquette variable is denoted by $U_\Box$, and
$\sigma^\alpha$ are the three Pauli matrices.
In fact, it is misleading to write the left-hand side of (\ref{standard})
as a time derivative of $N_{\rm CS}$ or any other fixed-time functional
of the fields: the object on the right-hand side of (\ref{standard}) is
not a total time derivative (TTD). This shortcoming of the lattice
topological charge density (TCD) is well known and related to the difficulty
in defining on a lattice an analog of the continuum $N_{\rm CS}$ with the
correct properties under gauge transformations \cite{mooremu}. Recently ASY
\cite{asy} pointed out that deviations of the lattice topological charge per
unit time from TTD may give rise to systematic errors in the transition rate.
For smooth field configurations (\ref{standard}) tends to the continuum
$\dot N_{\rm CS}$. However, thermal field configurations on the lattice are
not smooth, hence (\ref{standard}) does not approach a TTD in the continuum
limit.

The idea behind the cooling method is to replace the real-time trajectory
in the phase space (obtained by integrating the classical equations of motion)
by another path between the same endpoints, the one along which phase-space
configurations are predominantly smooth. Lattice topological charge, computed
along such a path using the naive definition (\ref{standard}), would suffer
much less from systematic errors than the one computed along the real-time
trajectory.
We shall first present the method for a wide class of Hamiltonian systems and
later discuss its application to topological transitions. A simplified 
version of the method was used in \cite{af,afetal} to study the time flow of 
eigenvalues of the (lattice) Dirac Hamiltonian in the presence of 
the time dependent gauge field generated by the real time, finite 
temperature simulations, and it was shown that the cooling strips off
the high energy modes, while it does not affect the (long wave length)
eigenvalue modes which 
cross zero and which are responsible for the change of baryon number. 
The results reported in \cite{af} provided the first proof that the 
sphaleron-like transitions observed in the real time lattice simulations
are indeed long wave length excitations and not lattice artifacts.   

Let the real-time evolution of a dynamical system be described by a
Hamiltonian depending on coordinates $q_i$ and momenta $p_i$ ($i=1,...N$):
$H=\sum_ip_i^2/2+V(q)$.
We introduce, along with the real time $t$, a cooling time
$\tau$, of which all the dynamical variables are functions:
$p_i=p_i(t,\tau)$, $q_i=q_i(t,\tau)$. The $t$ dynamics at $\tau=0$ is given by
$$\partial_t q_i(t,0)=p_i(t,0); \ \ \ \ \
\partial_t p_i(t,0)=-\partial_{q_i}V(q(t,0)),$$
while the evolution in the $\tau$ direction is an overdamped motion
(cooling):
\begin{equation}
\partial_{\tau} q_i(t,\tau)=-\partial_{q_i}V(q(t,\tau)).
\label{dqdtau}\end{equation}
If we insist that the real-time equations for the coordinates
$\partial_t q_i(t,\tau)=p_i(t,\tau)$ hold everywhere in the $t,\tau$ plane,
then the $\tau$ dependence of momenta, consistent with (\ref{dqdtau}), is
given by
\begin{equation}
\partial_{\tau} p_i(t,\tau)=-p_j(\tau)\partial^2_{q_i,q_j}V(q(t,\tau)).
\label{dpdtau}\end{equation}
It is instructive to consider the $\tau$ evolution described by
(\ref{dqdtau})-(\ref{dpdtau}) in the vicinity of a static solution
$\partial_{q_i}V=0$. The equations of motion (\ref{dqdtau})-(\ref{dpdtau}) can
be linearized and dynamical variables can be chosen as eigenmodes of these
linearized equations: $\partial_\tau q_i=-\omega^2_i q_i$,
$\partial_\tau p_i=-\omega^2_i p_i$. Thus
in the vicinity of a static solution cooling leads to an exponential decay
of stable eigenmodes and exponential growth of unstable ones. Moreover, the
rate of decay (growth) is especially rapid for high-frequency modes.

Suppose there is a coordinate (call it $Q$) along which $V$ is periodic. For
simplicity we can assume that variations of $Q$ are stable eigenmodes of the
linearized equations of motion in the vicinity of periodic minima of $V$
and the unstable ones in the vicinity of saddle points separating these minima.
For sufficiently large $\tau$ $Q(t,0)$ will be mapped by cooling to $Q(t,\tau)$
in a close vicinity of a minimum, unless $Q(t,0)$ lies close to a saddle point.
As a result, if the $Q(t,0)$ motion consists of transitions between minima
separated by fluctuations close to a minimum, the picture will sharpen under
cooling: fluctuations of $Q(t,\tau)$ near a minimum will have a smaller
magnitude, whereas transitions between vacua (the time spent in the vicinity of
a saddle point) will become shorter. If this indeed is the case for $Q$, two
lessons can be learned. On the one hand, the average rate of transitions
between minima for $Q(t,\tau)$ should obviously be independent of $\tau$. On
the other hand, care should be exercised if $t$ is discretized for numerical
purposes: for large $\tau$ the transition times may become shorter than the $t$
time step.

Consider now cooling as a method for approximating topological charge. Lattice
topological charge per unit time is not a TTD. For smooth field configurations
a lattice TCD can be expanded in powers of the lattice spacing, the leading
term in the expansion being the continuum TCD. For a thermal field configuration
this expansion is formal since in three dimensions thermal configurations are
not smooth. By cooling, however, these configurations can be made smooth enough
for the expansion to make sense. Then a lattice TCD will give rise to much
smaller deviations from a TTD. We are therefore led to the following method of
determining the topological charge. For any two instances of real time consider
evolution of lattice fields along a $\Pi$-shaped trajectory in the $t,\tau$
plain, with $\tau$ sufficiently large to give a smooth cooled configuration,
while, at the same time, $\tau\ll t$. Variation of topological charge with
$\tau$ is found by replacing $t$ derivatives with $\tau$ derivatives in the
expression for TCD, whereas $t$ derivatives at $\tau\neq 0$, required for the
horizontal part of the trajectory, are found using (\ref{dpdtau}). If the space
integral of a lattice TCD were a TTD, the topological charge along any such
trajectory would be independent of $\tau$. This is not the case for an
approximate lattice TCD. However, for sufficiently deep cooling the
topological charge becomes nearly independent of $\tau$. Indeed, deviations from
a TTD are suppressed along the horizontal part of the trajectory. The
vertical parts of the trajectory, where deviations from a TTD are not
suppressed, are of a fixed (short) duration and give rise to a small error in
determination of topological charge. Unlike in the $\tau=0$ case, this error
does not accumulate.
\begin{figure}
\begin{Large}
\setlength{\unitlength}{0.240900pt}
\ifx\plotpoint\undefined\newsavebox{\plotpoint}\fi
\sbox{\plotpoint}{\rule[-0.200pt]{0.400pt}{0.400pt}}%
\begin{picture}(1800,450)(0,0)
\font\gnuplot=cmr10 at 10pt
\gnuplot
\sbox{\plotpoint}{\rule[-0.200pt]{0.400pt}{0.400pt}}%
\put(878,44){\makebox(0,0){$t$}}
\put(106,247){\makebox(0,0)[l]{$\tau$}}
\put(192,89){\vector(0,1){158}}
\put(192,247){\line(0,1){158}}
\put(192,405){\vector(1,0){686}}
\put(878,405){\line(1,0){686}}
\put(1564,405){\vector(0,-1){158}}
\put(1564,247){\line(0,-1){158}}
\put(861,89){\vector(1,0){17}}
\put(192,89){\usebox{\plotpoint}}
\multiput(192,89)(20.756,0.000){67}{\usebox{\plotpoint}}
\put(1564,89){\usebox{\plotpoint}}
\sbox{\plotpoint}{\rule[-0.400pt]{0.800pt}{0.800pt}}%
\put(192,89){\raisebox{-.8pt}{\makebox(0,0){$\Diamond$}}}
\put(1564,89){\raisebox{-.8pt}{\makebox(0,0){$\Diamond$}}}
\end{picture}
\end{Large}
\caption{A $\Pi$-shaped trajectory in the $t,\tau$ plain used for 
determination of topological charge with cooling. The endpoints of a $\tau=0$
real-time trajectory, shown by a dotted line, are denoted by diamonds.}
\label{taut}
\end{figure}
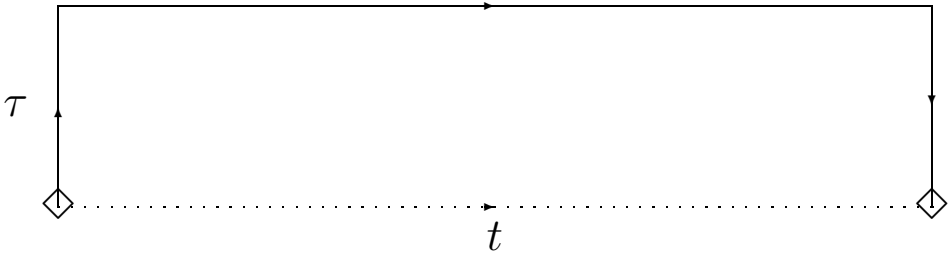
We now write down the cooling equations for the Yang-Mills theory described 
on the lattice by the Kogut-Susskind Hamiltonian
$$H_{YM}={1\over 2}\sum_{j,n} E^\alpha_{j,n} E^\alpha_{j,n}
+ \sum_\Box\left(1-{1\over 2}{\rm Tr}U_\Box\right).$$
In doing so it should be kept in mind
that link matrices $U$ are SU(2) group elements, each belonging to its own
group. Hence the cooling equation for a link $U$ can be written using
covariant derivatives $D^\alpha$ on the group to which $U$ belongs:
\begin{equation}
\partial_\tau U=-D^\alpha H_{YM} D^\alpha U.
\label{dudtau}\end{equation}
Here $\alpha$ is an adjoint SU(2) index.
If we choose to work with right covariant derivatives, 
$D^\alpha U=-iU\sigma^\alpha$.
Since $H_{YM}$ is linear in any given $U$, $D^\alpha H_{YM}$ is obtained
by retaining of $H_{YM}$ only terms proportional to $U$, with $U$ replaced
by $D^\alpha U$.

The cooling equations for the lattice electric fields $E^\alpha$ are derived, 
in analogy with (\ref{dpdtau}), by requiring, separately for each link, that the
Hamiltonian equation of motion ${\dot U}=-iU\sigma^\alpha E^\alpha$ holds
everywhere in the $t,\tau$ plane. Denoting $E\equiv\sigma^\alpha E^\alpha$
we then obtain
\begin{equation}
\partial_\tau E=i\partial_\tau(U^\dagger {\dot U})
=(\partial_\tau U^\dagger)U E+iU^\dagger\{H_{YM},\partial_\tau U\},
\label{dedtau}\end{equation}
where $\{\}$ are Poisson brackets and $\partial_\tau U$ is substituted from
(\ref{dudtau}) (we refrain here from writing the somewhat cumbersome explicit
expression for $\{H_{YM},\partial_\tau U\}$). 

The cooling method as described for the Yang-Mills theory applies essentially 
unchanged to the Yang-Mills-Higgs system whose Hamiltonian is
\begin{equation}
H_H=H_{YM}+\sum_j|\pi_j|^2+ \sum_{j,n}|\phi_{j+n}-U_{j,
n}^\dagger\phi_j|^2+ \lambda\sum_j\left(|\phi_j|^2-v^2\right)^2.
\label{hh}\end{equation}
Our notation for the scalar fields is standard, {\it i.e.,} to every lattice 
site $j$ we assign the scalar doublet $\phi_j$ along with its conjugate 
momentum $\pi_j$.
For the gauge fields one simply replaces $H_{YM}$ by $H_H$ in 
(\ref{dudtau})-(\ref{dedtau}). 
These equations are complemented by the cooling equations
for the scalar fields and momenta, which are a straightforward implementation
of (\ref{dqdtau})-(\ref{dpdtau}).

\section{The rate in the Yang-Mills-Higgs theory}\label{ymh}
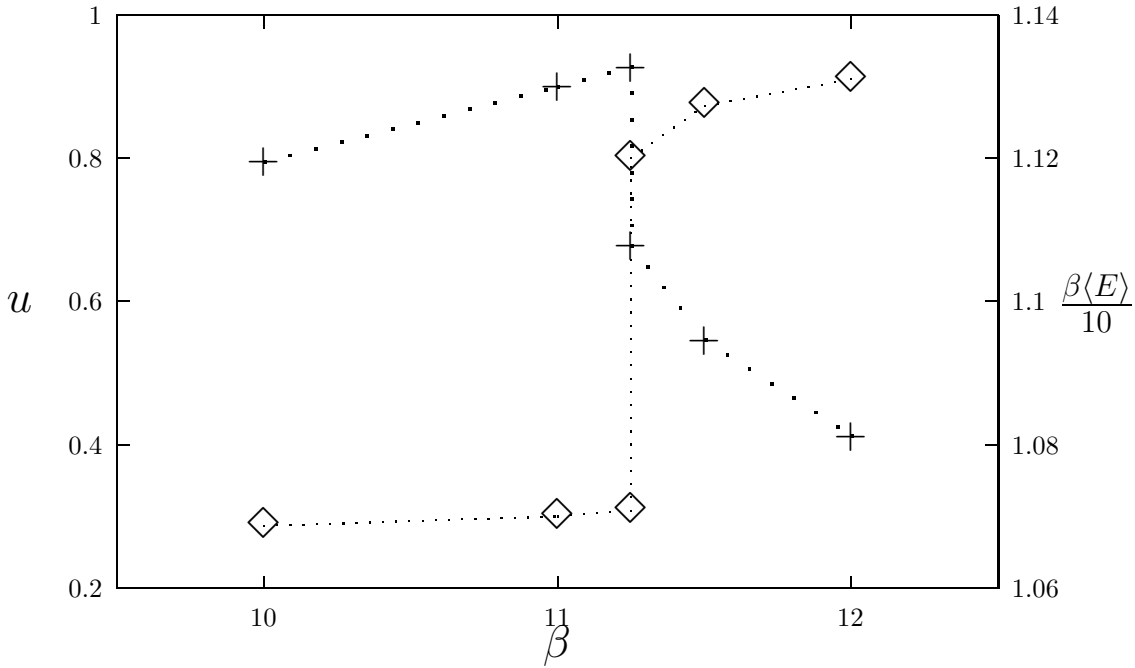
\begin{figure}[t]
\begin{Large}
\setlength{\unitlength}{0.240900pt}
\ifx\plotpoint\undefined\newsavebox{\plotpoint}\fi
\sbox{\plotpoint}{\rule[-0.200pt]{0.400pt}{0.400pt}}%
\begin{picture}(1800,1080)(0,0)
\font\gnuplot=cmr10 at 10pt
\gnuplot
\sbox{\plotpoint}{\rule[-0.200pt]{0.400pt}{0.400pt}}%
\put(197.0,134.0){\rule[-0.200pt]{4.818pt}{0.400pt}}
\put(175,134){\makebox(0,0)[r]{0.2}}
\put(1561.0,134.0){\rule[-0.200pt]{4.818pt}{0.400pt}}
\put(197.0,359.0){\rule[-0.200pt]{4.818pt}{0.400pt}}
\put(175,359){\makebox(0,0)[r]{0.4}}
\put(1561.0,359.0){\rule[-0.200pt]{4.818pt}{0.400pt}}
\put(197.0,585.0){\rule[-0.200pt]{4.818pt}{0.400pt}}
\put(175,585){\makebox(0,0)[r]{0.6}}
\put(1561.0,585.0){\rule[-0.200pt]{4.818pt}{0.400pt}}
\put(197.0,810.0){\rule[-0.200pt]{4.818pt}{0.400pt}}
\put(175,810){\makebox(0,0)[r]{0.8}}
\put(1561.0,810.0){\rule[-0.200pt]{4.818pt}{0.400pt}}
\put(197.0,1035.0){\rule[-0.200pt]{4.818pt}{0.400pt}}
\put(175,1035){\makebox(0,0)[r]{1}}
\put(1561.0,1035.0){\rule[-0.200pt]{4.818pt}{0.400pt}}
\put(428.0,134.0){\rule[-0.200pt]{0.400pt}{4.818pt}}
\put(428,89){\makebox(0,0){10}}
\put(428.0,1015.0){\rule[-0.200pt]{0.400pt}{4.818pt}}
\put(889.0,134.0){\rule[-0.200pt]{0.400pt}{4.818pt}}
\put(889,89){\makebox(0,0){11}}
\put(889.0,1015.0){\rule[-0.200pt]{0.400pt}{4.818pt}}
\put(1350.0,134.0){\rule[-0.200pt]{0.400pt}{4.818pt}}
\put(1350,89){\makebox(0,0){12}}
\put(1350.0,1015.0){\rule[-0.200pt]{0.400pt}{4.818pt}}
\put(1603,134){\makebox(0,0)[l]{1.06}}
\put(1561.0,134.0){\rule[-0.200pt]{4.818pt}{0.400pt}}
\put(1603,359){\makebox(0,0)[l]{1.08}}
\put(1561.0,359.0){\rule[-0.200pt]{4.818pt}{0.400pt}}
\put(1603,585){\makebox(0,0)[l]{1.1}}
\put(1561.0,585.0){\rule[-0.200pt]{4.818pt}{0.400pt}}
\put(1603,810){\makebox(0,0)[l]{1.12}}
\put(1561.0,810.0){\rule[-0.200pt]{4.818pt}{0.400pt}}
\put(1603,1035){\makebox(0,0)[l]{1.14}}
\put(1561.0,1035.0){\rule[-0.200pt]{4.818pt}{0.400pt}}
\put(197.0,134.0){\rule[-0.200pt]{333.406pt}{0.400pt}}
\put(1581.0,134.0){\rule[-0.200pt]{0.400pt}{217.051pt}}
\put(197.0,1035.0){\rule[-0.200pt]{333.406pt}{0.400pt}}
\put(45,584){\makebox(0,0){$u$}}
\put(1736,584){\makebox(0,0){${{\beta \langle E\rangle}\over 10}$}}
\put(889,44){\makebox(0,0){$\beta$}}
\put(197.0,134.0){\rule[-0.200pt]{0.400pt}{217.051pt}}
\put(428,232){\raisebox{-.8pt}{\makebox(0,0){$\Diamond$}}}
\put(889,247){\raisebox{-.8pt}{\makebox(0,0){$\Diamond$}}}
\put(1004,256){\raisebox{-.8pt}{\makebox(0,0){$\Diamond$}}}
\put(1004,810){\raisebox{-.8pt}{\makebox(0,0){$\Diamond$}}}
\put(1120,892){\raisebox{-.8pt}{\makebox(0,0){$\Diamond$}}}
\put(1350,934){\raisebox{-.8pt}{\makebox(0,0){$\Diamond$}}}
\put(428,232){\usebox{\plotpoint}}
\multiput(428,232)(20.745,0.675){23}{\usebox{\plotpoint}}
\multiput(889,247)(20.692,1.619){5}{\usebox{\plotpoint}}
\multiput(1004,256)(0.000,20.756){27}{\usebox{\plotpoint}}
\multiput(1004,810)(16.948,11.981){7}{\usebox{\plotpoint}}
\multiput(1120,892)(20.418,3.728){11}{\usebox{\plotpoint}}
\put(1350,934){\usebox{\plotpoint}}
\sbox{\plotpoint}{\rule[-0.400pt]{0.800pt}{0.800pt}}%
\put(428,804){\makebox(0,0){$+$}}
\put(889,922){\makebox(0,0){$+$}}
\put(1004,953){\makebox(0,0){$+$}}
\put(1004,672){\makebox(0,0){$+$}}
\put(1120,524){\makebox(0,0){$+$}}
\put(1350,373){\makebox(0,0){$+$}}
\sbox{\plotpoint}{\rule[-0.500pt]{1.000pt}{1.000pt}}%
\sbox{\plotpoint}{\rule[-0.600pt]{1.200pt}{1.200pt}}%
\sbox{\plotpoint}{\rule[-0.500pt]{1.000pt}{1.000pt}}%
\put(428,804){\usebox{\plotpoint}}
\multiput(428,804)(40.215,10.294){12}{\usebox{\plotpoint}}
\multiput(889,922)(40.080,10.804){3}{\usebox{\plotpoint}}
\multiput(1004,953)(0.000,-41.511){7}{\usebox{\plotpoint}}
\multiput(1004,672)(25.607,-32.671){4}{\usebox{\plotpoint}}
\multiput(1120,524)(34.701,-22.782){7}{\usebox{\plotpoint}}
\put(1350,373){\usebox{\plotpoint}}
\end{picture}
\end{Large}
\caption{The "string bit" order parameter (diamonds) and the average energy per 
degree of freedom normalized by the energy of a free theory (pluses) for the 
Yang-Mills-Higgs theory with the tree-level $m_H/m_W=0.423$. Note the 
discontinuity of both quantities at $\beta=11.25$. The error bars are smaller
than the plotting symbols. The dotted lines are to guide the eye.}
\label{stat12}
\end{figure}

We used the cooling method in order to study the real-time evolution of
Chern-Simons number in the SU(2) theory with and without the scalar field.
In (\ref{hh}) we chose $\lambda=0.09109$, corresponding to the tree-level 
$m_H/m_W$ ratio of 0.423.
Since the static properties of the classical theory closely resemble those of
the dimensionally reduced one, we expected our lattice system to undergo a 
strongly first order phase transition. This is indeed what we find: the system
exhibits a metastability at $beta=11.25$, with
an easily detectable latent heat and a large discontinuity in the conventionally
used "string bit" order parameter
$$u\equiv
\langle\phi^*_j U_{j,n}\phi_{j+n}/(|\phi^*_j||\phi_{j+n}|)^{1/2}\rangle$$
as shown in Figure \ref{stat12}. Therefore, according to the conventional 
wisdom \cite{rsrev}, the sphaleron regime should set in at temperatures 
immediately below the transition. At $\beta=12$ the sphaleron estimate for 
$\Gamma$ involves the Boltzmann factor $\exp(-\beta E_{\rm sph})$ which 
prohibits direct observation 
of sphaleron transitions in a feasible numerical simulation. 

\begin{figure}[t]
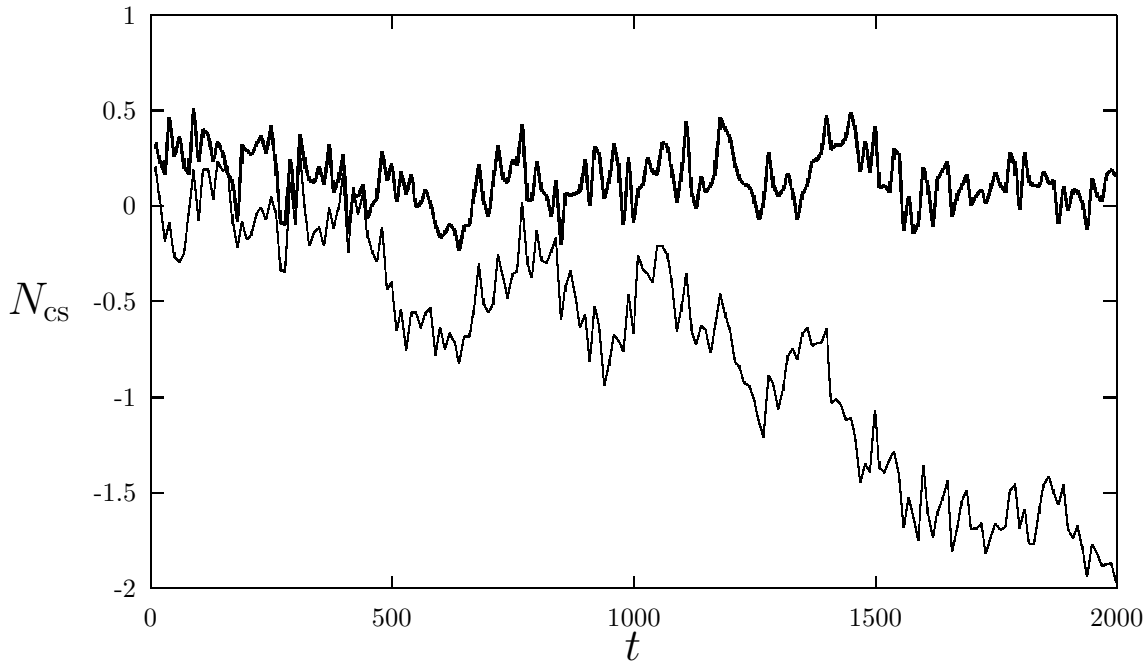

\begin{Large}
\setlength{\unitlength}{0.240900pt}
\ifx\plotpoint\undefined\newsavebox{\plotpoint}\fi
\sbox{\plotpoint}{\rule[-0.200pt]{0.400pt}{0.400pt}}%

\end{Large}
\caption{Time history of $N_{\rm CS}$ in the Yang-Mills-Higgs theory with
(the upper curve) and without (the lower curve) the cooling improvement.}
\label{cshiggs}
\end{figure}

Our numerical experiment at $\beta=12$ reveals a dramatic difference between
the naive and refined definitions of topological charge (Figure \ref{cshiggs}). 
While the former is a random walk with deviations as large as 2 for a
real-time trajectory of 2000 units, deviations of the latter never exceed 1.
In fact, the motion of the refined Chern-Simons number is completely consistent
with thermal fluctuations in the vicinity of one of the gauge-equivalent vacua
and involves no diffusion between the vacua. 
In order to eliminate a possibility
of diffusion with a time scale longer than 2000 units we followed the 
real-time 
evolution of the refined $N_{\rm CS}$ for $2\times 10^4$ time units. Again, we
detected
no sign of diffusion. 
Our results, together with observations of Ref. \cite{mt},
strongly suggest that the slow diffusion of the naively defined Chern-Simons
is spurious. Using the refined definition instead, we can place an upper bound
on the sphaleron transition rate, expressed in terms of the dimensionless 
coefficient $\kappa$: $\kappa=(\pi\beta)^4/(Vt) < 0.01$, where
$Vt$ is the total space-time volume of our simulation. 
Clearly, this upper bound cannot be used to confront the sphaleron 
estimate which it exceeds by many orders of magnitude.

\section{The rate in the Yang-Mills theory for small volumes}\label{ymtiny}
\begin{figure}[t]
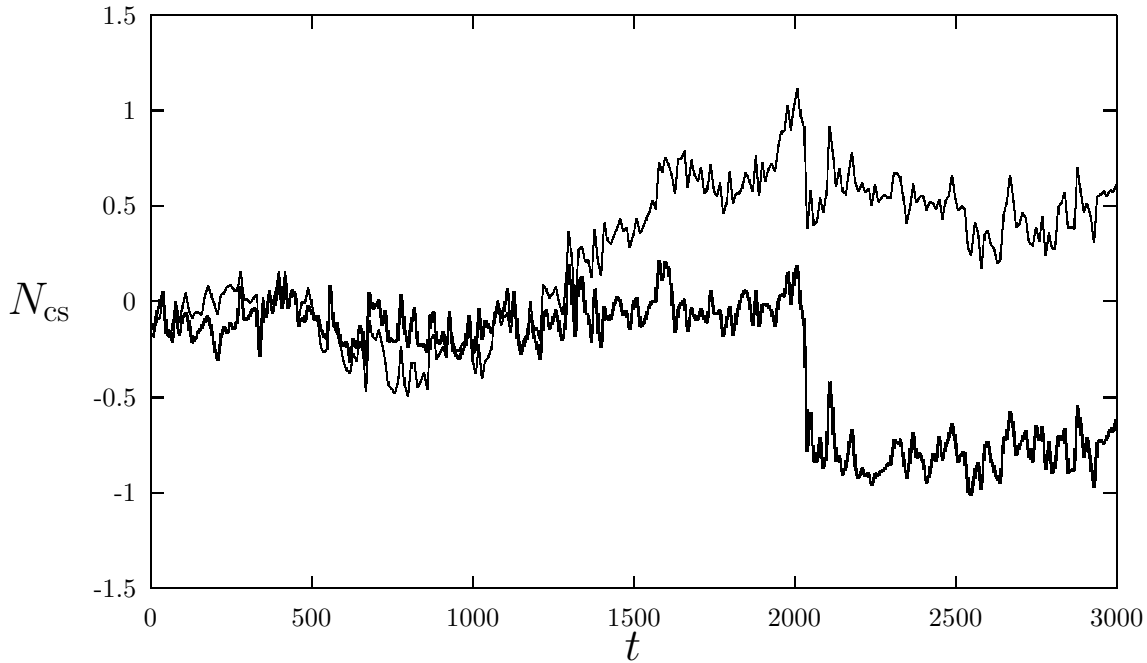

\begin{Large}
\setlength{\unitlength}{0.240900pt}
\ifx\plotpoint\undefined\newsavebox{\plotpoint}\fi
\sbox{\plotpoint}{\rule[-0.200pt]{0.400pt}{0.400pt}}%

\end{Large}
\caption{Time history of $N_{\rm CS}$ in the Yang-Mills
and without the cooling improvement at $\beta=L=12$. Both curves exhibit
rapid fluctuations and a sphaleron transition at $t\approx 2000$, but, for the
naive definition of $N_{\rm CS}$, there also is a slow diffusion upwards.}
\label{cs1212}
\end{figure}

Since we suspect the naive definition of topological charge density to give 
rise to systematic errors in $\Gamma$, it is necessary to revise the results of
\cite{su2rate} for the Yang-Mills theory, obtained using that definition. While 
our ultimate goal is to determine $\Gamma$ for lattice systems large enough
to rule out finite-size effects, it is instructive to first consider 
smaller-volume systems. As we observed in \cite{su2rate}, finite-size effects
in $\Gamma$ set in as soon as the linear size $L$ of the lattice is close to the
inverse temperature $\beta$. This phenomenon can be interpreted as follows. 
In the infinite-volume system there is no potential barrier separating
topologically distinct vacua. Indeed, such a barrier is prohibited by the scale
invariance of the classical theory. A finite size breaks the scale invariance 
and gives rise to the minimal barrier height of $18.15/L$ for the periodic
boundary conditions used here \cite{baaldas}.
Making use of this property, we can suitably adjust the $\beta/L$ ratio in order
to study the $N_{\rm CS}$ behavior in the sphaleron regime. With this goal in
mind, we performed a simulation at $\beta/L=1$. As Figure \ref{cs1212} shows, 
motion of the naively defined $N_{\rm CS}$ can be seen as consisting of three 
components: thermal fluctuations, rapid transitions with a magnitude close to
1, and slow diffusion. If the refined definition is used, the slow diffusion
component is absent from the $N_{\rm CS}$ time history. If we put this result
together with our findings for the Yang-Mills-Higgs system, a consistent 
picture emerges: the true sphaleron transitions occur rapidly, 
while the slow diffusion of the naive $N_{\rm CS}$ is a discretization
artifact. 

Next, we determine the transition rate $\Gamma$ in the small-volume case. 
This is most easily and efficiently done by counting the clearly distinguishable
sphaleron transitions. The rate is then the number of transitions per unit
volume per unit time. The advantage of this procedure over determining the
rate as the $N_{\rm CS}$ diffusion constant is that the sphaleron event count 
is the same with either $N_{\rm CS}$ measurement method. We therefore use the
naive definition of $N_{\rm CS}$ in this part of the work.

We have observed between 50 and 100 events for $\beta=L=10,12,14,16,20$.
Since any two consecutive events are widely
separated in time and unlikely to be correlated, we determine the rate with
the relative accuracy $n(\beta)^{-1/2}$, where $n(\beta)$ the total number of
observed events for a given $\beta$.
Our findings, summarized in Table \ref{bl1}, clearly show 
dependence of the effective $\kappa$ on $\beta$, and hence on the lattice
spacing. As can be seen in Figure \ref{beqlk}, for $10\leq\beta\leq 16$ this 
dependence is consistent with
$\kappa\propto\beta^{-1}$, but this simple proportionality cannot
accommodate the $\beta=20$ point in a convincing way.

Clearly, our small-volume measurements fail to
demonstrate the existence of continuum limit for the classical rate. 
Our simulations give support to ASY prediction $\kappa \sim 1/\beta$ 
except maybe for the largest value of $\beta$. However, as we shall see in the
following, a closer examination of the gauge field dynamics leaves room for
alternative explanations of this discretization dependence of the small-volume
rate. 
\begin{table}
\centerline{\begin{tabular}{|c|rrrrr|} \hline
{$\beta=L$} & {10} & {12} & {14} & {16} & {20} \\
\hline
{events} & {111} & {109} & {113} & {73} & {50} \\
\hline
{time} & {$5.55\times 10^5$} & {$8.2\times 10^5$} & {$1.23\times 10^6$}
& {$1.02\times 10^6$} & {$1.44\times 10^6$} \\
\hline
{$kappa$} & {$0.195\pm 0.018$} & {$0.155\pm 0.015$} & {$0.125\pm 0.012$} &
{$0.112\pm 0.013$} & {$0.0676\pm 0.0096$} \\
\hline
\end{tabular}}
\caption{Effective $\kappa$ for $\beta/L=1$.}
\label{bl1}
\end{table}
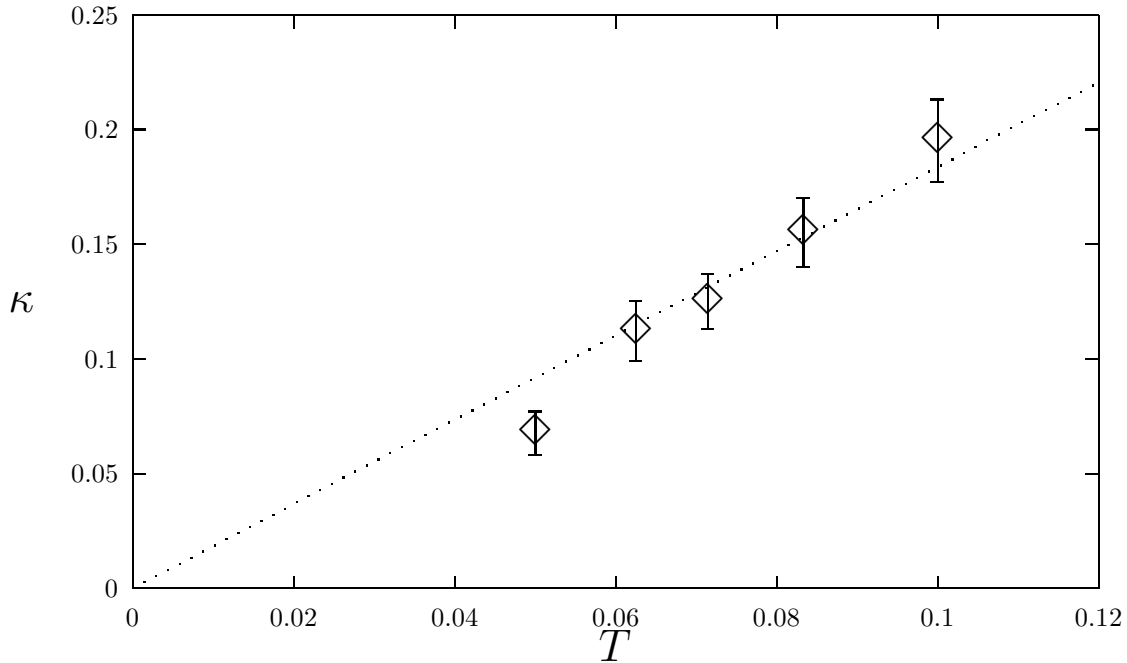
\begin{figure}[t]
\begin{Large}
\setlength{\unitlength}{0.240900pt}
\ifx\plotpoint\undefined\newsavebox{\plotpoint}\fi
\sbox{\plotpoint}{\rule[-0.200pt]{0.400pt}{0.400pt}}%
\begin{picture}(1800,1080)(0,0)
\font\gnuplot=cmr10 at 10pt
\gnuplot
\sbox{\plotpoint}{\rule[-0.200pt]{0.400pt}{0.400pt}}%
\put(219.0,134.0){\rule[-0.200pt]{4.818pt}{0.400pt}}
\put(197,134){\makebox(0,0)[r]{0}}
\put(1716.0,134.0){\rule[-0.200pt]{4.818pt}{0.400pt}}
\put(219.0,314.0){\rule[-0.200pt]{4.818pt}{0.400pt}}
\put(197,314){\makebox(0,0)[r]{0.05}}
\put(1716.0,314.0){\rule[-0.200pt]{4.818pt}{0.400pt}}
\put(219.0,494.0){\rule[-0.200pt]{4.818pt}{0.400pt}}
\put(197,494){\makebox(0,0)[r]{0.1}}
\put(1716.0,494.0){\rule[-0.200pt]{4.818pt}{0.400pt}}
\put(219.0,675.0){\rule[-0.200pt]{4.818pt}{0.400pt}}
\put(197,675){\makebox(0,0)[r]{0.15}}
\put(1716.0,675.0){\rule[-0.200pt]{4.818pt}{0.400pt}}
\put(219.0,855.0){\rule[-0.200pt]{4.818pt}{0.400pt}}
\put(197,855){\makebox(0,0)[r]{0.2}}
\put(1716.0,855.0){\rule[-0.200pt]{4.818pt}{0.400pt}}
\put(219.0,1035.0){\rule[-0.200pt]{4.818pt}{0.400pt}}
\put(197,1035){\makebox(0,0)[r]{0.25}}
\put(1716.0,1035.0){\rule[-0.200pt]{4.818pt}{0.400pt}}
\put(219.0,134.0){\rule[-0.200pt]{0.400pt}{4.818pt}}
\put(219,89){\makebox(0,0){0}}
\put(219.0,1015.0){\rule[-0.200pt]{0.400pt}{4.818pt}}
\put(472.0,134.0){\rule[-0.200pt]{0.400pt}{4.818pt}}
\put(472,89){\makebox(0,0){0.02}}
\put(472.0,1015.0){\rule[-0.200pt]{0.400pt}{4.818pt}}
\put(725.0,134.0){\rule[-0.200pt]{0.400pt}{4.818pt}}
\put(725,89){\makebox(0,0){0.04}}
\put(725.0,1015.0){\rule[-0.200pt]{0.400pt}{4.818pt}}
\put(978.0,134.0){\rule[-0.200pt]{0.400pt}{4.818pt}}
\put(978,89){\makebox(0,0){0.06}}
\put(978.0,1015.0){\rule[-0.200pt]{0.400pt}{4.818pt}}
\put(1230.0,134.0){\rule[-0.200pt]{0.400pt}{4.818pt}}
\put(1230,89){\makebox(0,0){0.08}}
\put(1230.0,1015.0){\rule[-0.200pt]{0.400pt}{4.818pt}}
\put(1483.0,134.0){\rule[-0.200pt]{0.400pt}{4.818pt}}
\put(1483,89){\makebox(0,0){0.1}}
\put(1483.0,1015.0){\rule[-0.200pt]{0.400pt}{4.818pt}}
\put(1736.0,134.0){\rule[-0.200pt]{0.400pt}{4.818pt}}
\put(1736,89){\makebox(0,0){0.12}}
\put(1736.0,1015.0){\rule[-0.200pt]{0.400pt}{4.818pt}}
\put(219.0,134.0){\rule[-0.200pt]{365.445pt}{0.400pt}}
\put(1736.0,134.0){\rule[-0.200pt]{0.400pt}{217.051pt}}
\put(219.0,1035.0){\rule[-0.200pt]{365.445pt}{0.400pt}}
\put(45,584){\makebox(0,0){$\kappa$}}
\put(977,44){\makebox(0,0){$T$}}
\put(219.0,134.0){\rule[-0.200pt]{0.400pt}{217.051pt}}
\put(851,378){\raisebox{-.8pt}{\makebox(0,0){$\Diamond$}}}
\put(1009,538){\raisebox{-.8pt}{\makebox(0,0){$\Diamond$}}}
\put(1122,585){\raisebox{-.8pt}{\makebox(0,0){$\Diamond$}}}
\put(1272,693){\raisebox{-.8pt}{\makebox(0,0){$\Diamond$}}}
\put(1483,837){\raisebox{-.8pt}{\makebox(0,0){$\Diamond$}}}
\put(851.0,343.0){\rule[-0.200pt]{0.400pt}{16.622pt}}
\put(841.0,343.0){\rule[-0.200pt]{4.818pt}{0.400pt}}
\put(841.0,412.0){\rule[-0.200pt]{4.818pt}{0.400pt}}
\put(1009.0,491.0){\rule[-0.200pt]{0.400pt}{22.645pt}}
\put(999.0,491.0){\rule[-0.200pt]{4.818pt}{0.400pt}}
\put(999.0,585.0){\rule[-0.200pt]{4.818pt}{0.400pt}}
\put(1122.0,541.0){\rule[-0.200pt]{0.400pt}{20.958pt}}
\put(1112.0,541.0){\rule[-0.200pt]{4.818pt}{0.400pt}}
\put(1112.0,628.0){\rule[-0.200pt]{4.818pt}{0.400pt}}
\put(1272.0,639.0){\rule[-0.200pt]{0.400pt}{26.017pt}}
\put(1262.0,639.0){\rule[-0.200pt]{4.818pt}{0.400pt}}
\put(1262.0,747.0){\rule[-0.200pt]{4.818pt}{0.400pt}}
\put(1483.0,772.0){\rule[-0.200pt]{0.400pt}{31.317pt}}
\put(1473.0,772.0){\rule[-0.200pt]{4.818pt}{0.400pt}}
\put(1473.0,902.0){\rule[-0.200pt]{4.818pt}{0.400pt}}
\put(219,134){\usebox{\plotpoint}}
\put(219.00,134.00){\usebox{\plotpoint}}
\put(237.36,143.68){\usebox{\plotpoint}}
\put(255.84,153.12){\usebox{\plotpoint}}
\put(274.16,162.88){\usebox{\plotpoint}}
\put(292.64,172.32){\usebox{\plotpoint}}
\put(311.00,182.00){\usebox{\plotpoint}}
\put(329.36,191.68){\usebox{\plotpoint}}
\put(347.84,201.12){\usebox{\plotpoint}}
\put(366.16,210.88){\usebox{\plotpoint}}
\put(384.64,220.32){\usebox{\plotpoint}}
\put(403.00,230.00){\usebox{\plotpoint}}
\put(421.36,239.68){\usebox{\plotpoint}}
\put(439.84,249.12){\usebox{\plotpoint}}
\put(458.16,258.88){\usebox{\plotpoint}}
\put(476.12,269.27){\usebox{\plotpoint}}
\put(494.56,278.78){\usebox{\plotpoint}}
\put(512.88,288.54){\usebox{\plotpoint}}
\put(531.28,298.14){\usebox{\plotpoint}}
\put(549.72,307.65){\usebox{\plotpoint}}
\put(568.04,317.42){\usebox{\plotpoint}}
\put(586.56,326.78){\usebox{\plotpoint}}
\put(604.88,336.54){\usebox{\plotpoint}}
\put(623.28,346.14){\usebox{\plotpoint}}
\put(641.72,355.65){\usebox{\plotpoint}}
\put(660.04,365.42){\usebox{\plotpoint}}
\put(678.56,374.78){\usebox{\plotpoint}}
\put(696.88,384.54){\usebox{\plotpoint}}
\put(715.28,394.14){\usebox{\plotpoint}}
\put(733.72,403.65){\usebox{\plotpoint}}
\put(752.04,413.42){\usebox{\plotpoint}}
\put(770.56,422.78){\usebox{\plotpoint}}
\put(788.88,432.54){\usebox{\plotpoint}}
\put(807.28,442.14){\usebox{\plotpoint}}
\put(825.72,451.65){\usebox{\plotpoint}}
\put(844.04,461.42){\usebox{\plotpoint}}
\put(862.56,470.78){\usebox{\plotpoint}}
\put(880.88,480.54){\usebox{\plotpoint}}
\put(899.28,490.14){\usebox{\plotpoint}}
\put(917.72,499.65){\usebox{\plotpoint}}
\put(936.04,509.42){\usebox{\plotpoint}}
\put(954.56,518.78){\usebox{\plotpoint}}
\put(972.88,528.54){\usebox{\plotpoint}}
\put(991.02,538.61){\usebox{\plotpoint}}
\put(1009.20,548.60){\usebox{\plotpoint}}
\put(1027.60,558.19){\usebox{\plotpoint}}
\put(1045.92,567.96){\usebox{\plotpoint}}
\put(1064.45,577.31){\usebox{\plotpoint}}
\put(1082.76,587.07){\usebox{\plotpoint}}
\put(1101.20,596.60){\usebox{\plotpoint}}
\put(1119.60,606.19){\usebox{\plotpoint}}
\put(1137.92,615.96){\usebox{\plotpoint}}
\put(1156.45,625.31){\usebox{\plotpoint}}
\put(1174.76,635.07){\usebox{\plotpoint}}
\put(1193.20,644.60){\usebox{\plotpoint}}
\put(1211.60,654.19){\usebox{\plotpoint}}
\put(1229.92,663.96){\usebox{\plotpoint}}
\put(1248.45,673.31){\usebox{\plotpoint}}
\put(1266.76,683.07){\usebox{\plotpoint}}
\put(1285.20,692.60){\usebox{\plotpoint}}
\put(1303.61,702.19){\usebox{\plotpoint}}
\put(1321.92,711.96){\usebox{\plotpoint}}
\put(1340.45,721.31){\usebox{\plotpoint}}
\put(1358.76,731.07){\usebox{\plotpoint}}
\put(1377.20,740.60){\usebox{\plotpoint}}
\put(1395.61,750.19){\usebox{\plotpoint}}
\put(1413.92,759.96){\usebox{\plotpoint}}
\put(1432.45,769.31){\usebox{\plotpoint}}
\put(1450.76,779.07){\usebox{\plotpoint}}
\put(1469.20,788.60){\usebox{\plotpoint}}
\put(1487.61,798.19){\usebox{\plotpoint}}
\put(1505.92,807.96){\usebox{\plotpoint}}
\put(1523.84,818.42){\usebox{\plotpoint}}
\put(1542.33,827.84){\usebox{\plotpoint}}
\put(1560.64,837.61){\usebox{\plotpoint}}
\put(1579.12,847.06){\usebox{\plotpoint}}
\put(1597.49,856.73){\usebox{\plotpoint}}
\put(1615.84,866.42){\usebox{\plotpoint}}
\put(1634.33,875.84){\usebox{\plotpoint}}
\put(1652.64,885.61){\usebox{\plotpoint}}
\put(1671.12,895.06){\usebox{\plotpoint}}
\put(1689.49,904.73){\usebox{\plotpoint}}
\put(1707.84,914.42){\usebox{\plotpoint}}
\put(1726.33,923.84){\usebox{\plotpoint}}
\put(1736,929){\usebox{\plotpoint}}
\end{picture}
\end{Large}
\caption{Effective $\kappa$ as a function of $T\equiv 1/\beta$ in the 
Yang-Mills
theory for $\beta=L$. The dotted line is a linear fit through the origin 
using all the data except the $\beta=20$ point.}
\label{beqlk}
\end{figure}
\section{The large-volume rate in the Yang-Mills theory}\label{ymbig}
In order to eliminate the systematic error in $\Gamma$ resulting from the naive
definition of topological charge we performed a series of large-volume
simulations similar to the one reported in \cite{su2rate}, but this time using 
the refined definition of $N_{\rm CS}$.
The cooling method is very costly computationally. For this
reason we were unable to produce a measurement sample of the same size as in
\cite{su2rate}, with a comparable numerical effort. Our new results therefore
contain larger statistical errors. In order to make use of the existing data
from \cite{su2rate} we adopted the following procedure.
Whatever definition of $N_{\rm CS}$ is used, the rate $\Gamma$ is found by 
fitting the average squared topological charge $\langle B^2(t)\rangle$ to a
linear function of the time interval $t$ in a suitable range of values of $t$.
The rate is then the slope of that linear function. To rephrase, the rate can
be determined given a numerical estimate for 
the joint probability distribution of
$D_i\equiv\langle B^2(t_i)\rangle$, where $t_i$ 
($1\leq i\leq l$) is a discrete
set of $t$ values to be used in the fit. We assume this probability distribution
to have the normal form
\begin{equation}
P(\{D\})\propto\exp\left(-{1\over 2}(D_i-{\overline D_i})(C^{-1})_{ij}
(D_i-{\overline D_i})\right),\label{pd}\end{equation}
where $C$ is the estimated covariance matrix and ${\overline D_i}$ is the
estimate of $D_i$ obtained by averaging over the data sample
({\it i.e.,} ${\overline D_i}\rightarrow D_i$ with increasing measurement 
accuracy \cite{steve}). We shall take (\ref{pd}) to represent results of our
high-statistics study \cite{su2rate}
In the current simulation we measure the improved topological
charge $B'(t)$ along with the naive one $B(t)$.  We then find the corresponding
${\overline D_i}$ and ${\overline D'_i}$, which we can jointly call
${\overline\Delta_j}$, $1\leq i\leq 2l$, meaning that 
${\overline\Delta_j}\equiv{\overline D_j}$  for $j\leq l$ and that 
${\overline\Delta_j}\equiv{\overline D'_{j+l}}$  for $j>l$.
We also estimate the covariance matrix ${\cal C}$ for $\{{\overline\Delta}\}$,
leading to the joint probability distribution $\Pi$ for $\{\Delta\}$ completely
analogous to (\ref{pd}):
\begin{equation}
\Pi(\{D\},\{D'\})\propto\exp\left(-{1\over 2}(\Delta_i-{\overline \Delta_i})
({\cal C}^{-1})_{ij}
(\Delta_i-{\overline \Delta_i})\right),\label{pd1}\end{equation}
It is obvious that for a fixed set of $\{D\}$
values of $\{D'\}$, averaged with the corresponding conditional probability,
will be shifted away from $\{{\overline D'}\}$ appearing in (\ref{pd1}). 
This also holds true if the distribution (\ref{pd}) of $\{D\}$ is given instead
of a sharp set of values. We then average $\Pi$ over $\{D\}$ with the
weight $P$ to obtain the corrected distribution for $\{D'\}$:
\begin{equation}
P'(\{D'\})=\int[dD]\Pi(\{D\},\{D'\})P(\{D\}).\label{piprime}\end{equation}
We refrain from writing down the somewhat cumbersome result of the trivial
Gaussian integration (\ref{piprime}). It is clear that $P'(\{D'\})$ is again
a normal distribution, with corrections to both averages of $D'_i$ and to 
their covariance matrix.

We now present results of our analysis. Table \ref{bigvol} summarizes the
large-volume simulations we performed. Throughout the simulations the
Hamiltonian time step was 0.05, and 5000 thermalization steps \cite{thalgs}
were performed before every Hamiltonian trajectory. We used cooling depth
$\tau=0.2$ for the improved determination of $N_{\rm CS}$. Within our 
measurement accuracy further cooling made no difference in the rate.
As in our previous work,
squared topological charge for a given time interval was first averaged 
within every Hamiltonian trajectory, and these averages, assumed independent,
served as an input for further statistical analysis, as described. While
the dimensionless coefficient $\kappa$ tends to decrease as a function of beta,
our results cannot be said to favor the $1/\beta$ behavior of $\kappa$ 
advocated in Ref. \cite{asy}. 
\begin{table}
\centerline{
\begin{tabular}{|ccccc|} \hline
{$\beta$} & {$L$} & {length $\times$ number of trajectories} & {$\kappa$} &
{biased $\kappa$} \\
\hline
{10} & {32} & {2000 $\times$ 20} & {1.48$\pm$0.08} & {1.44$\pm$0.03} \\
{12} & {32} & {2000 $\times$ 20} & {1.36$\pm$0.09} & {1.39$\pm$0.06} \\
{14} & {32} & {2000 $\times$ 10} & {1.33$\pm$0.12} & {1.38$\pm$0.10} \\
{14} & {36} & {1500 $\times$ 20} & {1.27$\pm$0.15} & {} \\
{16} & {36} & {1500 $\times$ 20} & {1.28$\pm$0.10} & {} \\
\hline
\end{tabular}}
\caption{Summary of the large-volume study of $\kappa$ in the Yang-Mills 
theory.}
\label{bigvol}
\end{table}
\begin{figure}[thb]
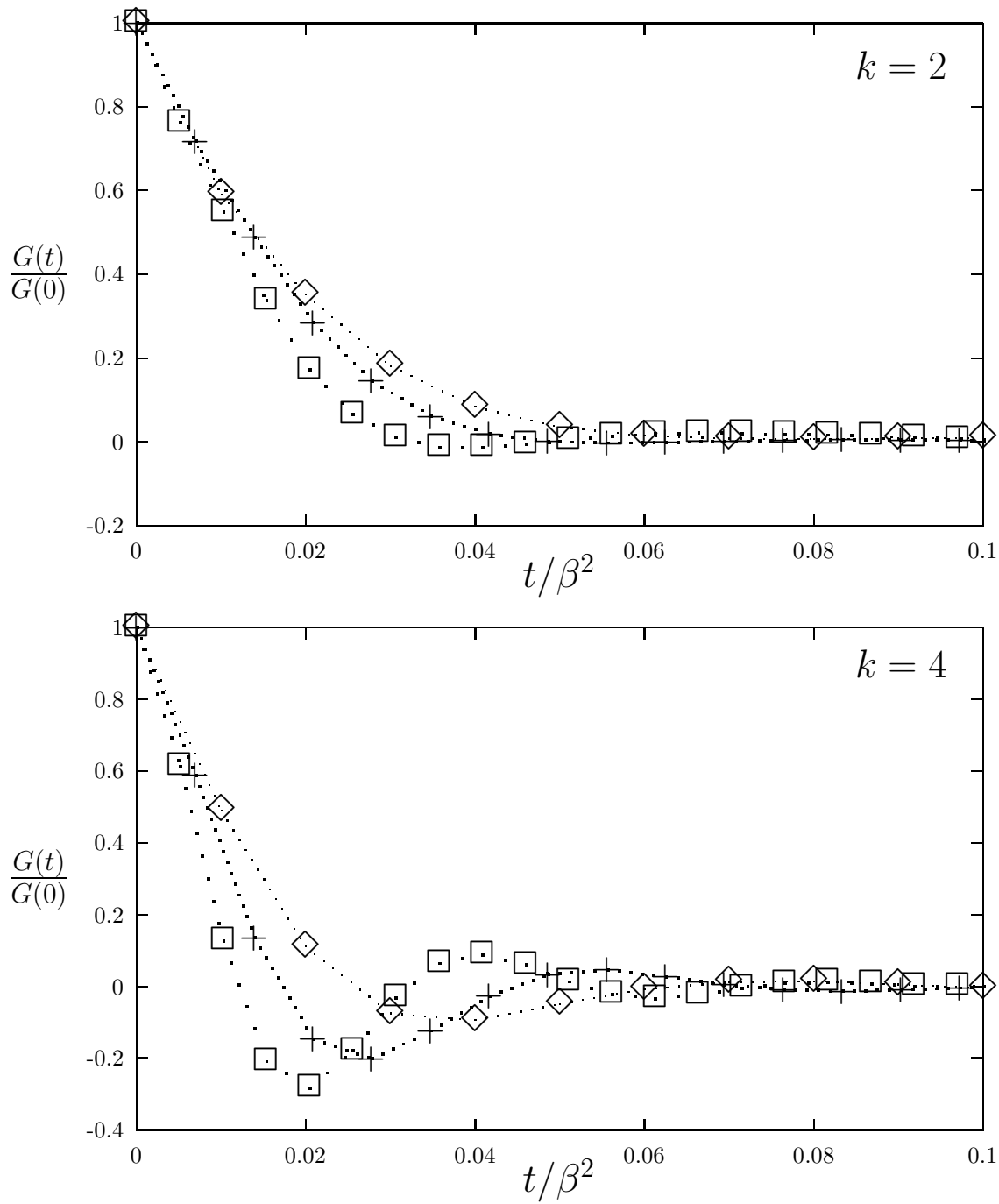

\begin{Large}
\setlength{\unitlength}{0.240900pt}
\ifx\plotpoint\undefined\newsavebox{\plotpoint}\fi
\sbox{\plotpoint}{\rule[-0.200pt]{0.400pt}{0.400pt}}%
%
\end{Large}
\caption{Real-time correlation functions $G(k,t)$ of gauge fields in the
Coulomb gauge plotted against $t/\beta^2$. The value of $k$ is indicated on
each plot. Each $G(k,t)$ is normalized by its respective value at $t=0$.
Diamonds correspond to $\beta=10$, plus signs to $\beta=12$ and squares to 
$\beta=14$ for a $32^3$ lattice. In addition, the open circles in the $k=0$ 
plot show $G(0,t)/G(0,0)$ for $\beta=14$ on a $14^3$ lattice. Wherever the
errorbars are not shown, they are smaller than the plotting symbols.}
\label{dres}
\end{figure}
\section{Dynamics of gauge fields in the Coulomb gauge}\label{glue}
We have seen in both large and small volumes the tendency of the rate to
decrease as the continuum limit is approached. A decrease of $\kappa$ as a 
function of $\beta$ was advocated in Ref. \cite{asy}. 
In a nutshell, Ref. \cite{asy} argued as follows. Baryon-number violating
processes mainly involve gauge fields with wavelengths $\sim\beta$, whose
motion is overdamped due to interaction with hard modes. This overdamping
leads to a $\beta^{-1}$ behavior of $\kappa$. While in the small-volume case
we observed this behavior of $\kappa$ in a range of values of $\beta$, the
decrease in $\kappa$ for large volumes was significantly slower than 
$\sim 1/\beta$. We therefore thought it useful to
study the motion of the long-wavelength gauge fields in some detail.

Gauge fields can only be defined by choice of a gauge. In this work we chose
the Coulomb gauge: on the one hand, it is often used in the relevant analytical
work \cite{brpis}; on the other hand, lattice gauge-fixing techniques have 
been developed for this gauge. The lattice Coulomb gauge condition is usually 
defined for every 
lattice site $j$ in terms of $A_{j,n}\equiv(1/(2i)(U_{j,n}-U^\dagger_{j,n})$:
\begin{equation}
\partial\dot A_j\equiv\sum_n(A_{j,n}-A_{j-n,n})=0,
\label{coug}\end{equation}
where $n$ denotes the three lattice directions. For the SU(2) group this 
condition can be obtained by extremizing 
\begin{equation}
{\rm Tr}\sum_{j,n}U_{j,n} 
\label{sumtr}\end{equation}
with respect to gauge transformations. This gauge is known to lead to Gribov 
copies; the unwanted copies are discarded by requiring the absolute maximum of
(\ref{sumtr}). It is important to note that in a system with periodic boundary
conditions (\ref{coug}) and (\ref{sumtr}) are preserved by global gauge
transformations. This additional gauge freedom is not important as long as
we only consider static properties of gauge fields, {\it e.g.,} screening
correlation functions, but it must be removed if we wish to compare gauge
fields at different instances of dynamical evolution. We do so by imposing two
additional global conditions: 
$${\rm Tr}\sigma^1\sum_jA_{j,1} ={\rm Tr}\sigma^2\sum_jA_{j,1}
={\rm Tr}\sigma^1\sum_jA_{j,2}=0.$$

In order to fix the gauge, we employed a combination of iterative overrelaxation
and Fourier acceleration algorithms. The former was used to attain the accuracy
of $10^{-3}$ for ${\rm Tr}(\partial\dot A)^2$ per site, and the latter to reduce
${\rm Tr}(\partial\dot A)^2$ to below $10^{-6}$. We took Gribov copies into 
account by repeatedly performing first a random gauge transformation, then the 
Coulomb gauge fixing on the same field configuration, and choosing among the 
resulting gauge copies the one with the largest ${\rm Tr}\sum_{j,n}U_{j,n}$. 
Within our measurement accuracy, this procedure had no effect on the 
observables.

We first determined the magnetic screening length from screening correlation
functions of the form
\begin{equation}
G(z)\equiv{\rm Tr}\sum_{z_0}\langle{\overline A}(z_0)\cdot{\overline A}(z_0+z)
\rangle, \label{screen}\end{equation}
where ${\overline A}(z_0)$ is the average of $A$ over the plane with one lattice
coordinate fixed to $z_0$. We further average $G(z)$ over the three directions
of the displacement $z$ (we will use the same notation for the average over
directions). The magnetic mass is found by fitting $G(z)$ to 
$C\exp(-mL/2)\cosh(m(z-L/2))$. The results for $\beta=10,12,14$ and $L=32$,
tabulated in Table \ref{mmag}, are accurately described by 
$m=(2.630\pm 0.005)/\beta$ in lattice units, {\it i.e.,} 
$m=(0.657\pm 0.001)g^2T$ in the continuum notation. Thus we find that in the
classical theory and in the range of temperatures considered $m$ increases
with $g^2T$ steeper than $0.466g^2T$ reported in Ref. \cite{rank} for the
high-temperature quantum theory. In that work correlation functions were 
computed at much lower $\beta\leq 3.47$ and in Landau, rather than Coulomb,
gauge. Note, however, that both in Ref. \cite{rank} and in the current work
$G(z)$ could only be described by $C\exp(-mL/2)\cosh(m(z-L/2))$ for $z$ close
to $L/2$, whereas for small values of $z$ the decay of $G(z)$ is noticeably
slower. That is to say, $1/m$ underestimates distances for which transverse
gauge fields are strongly correlated in space.

Similarly, we estimated the Debye screening mass $m_D$ from
\begin{equation}
G_D(z)\equiv{\rm Tr}\sum_{z_0}\langle{\overline E_z}(z_0){\overline E_z}(z_0+z)
\rangle, \label{deb}\end{equation}
where, {\it e.g.,}  ${\overline E_z}(z_0)$ denotes the average of the 
Cartesian $z$ component of the lattice color electric field over the lattice 
plane whose $z$ coordinate is $z_0$. We further average $G_D$ over the three
lattice directions and fit the result to $C\exp(-m_DL/2)\cosh(m_D(z-L/2))$,
just as we did for the magnetic mass. The results are summarized in the
bottom line of Table \ref{mmag}. Assuming $m_D\propto 1/\sqrt{\beta}$, our data
is best described by $m_D=(2.328\pm 0.002)/\sqrt{\beta}$, slightly above the
perturbative estimate $m_D=2.01/\sqrt{\beta}$ \cite{arnonly}. Therefore 
for $\beta\leq 14$ $m_D/m\leq 3.7$
in the classical theory, and these two scales are not completely
separated. This point could be important for understanding our results
for the rate.
\begin{table}
\centerline{\begin{tabular}{|cccc|} \hline
{$\beta$} & {10} & {12} & {14} \\
{$m$} & {$0.27\pm 0.02$} & {$0.23\pm 0.01$} & {$0.184\pm 0.006$} \\
{$m_D$} & {$0.712\pm 0.025$} & {$0.69\pm 0.04$} & {$0.68\pm 0.04$} \\
\hline
\end{tabular}}
\caption{Magnetic and Debye screening masses for the $32^3$ classical system.}
\label{mmag}
\end{table}

Next, we consider the dynamical behavior of gauge fields in the Coulomb
gauge. We determined real-time correlation functions of fields with a 
definite latice momentum $2\pi k/L$ in one of the three lattice directions, 
for $0\leq k\leq 4$. More specifically, we studied
$$G(k,t)\equiv{\rm Re}\, \langle A^*_{k,p}(t)A_{k,p}(0)\rangle,$$
where $p$ runs over the two transverse polarizations. We further averaged 
$G(k,t)$ over the three directions of $k$, and the notation $G(k,t)$ is
from now on used for that average. The resulting correlation 
functions, plotted in Figure \ref{dres}, exhibit the following important 
features.
First the motion of the gauge fields is clearly overdamped for $k=0,1$: the 
corresponding $G(k,t)$ decay without oscillations. For $k=3,4$ the field motion
is underdamped, and an oscillatory behavior of $G(k,t)$ is clearly observed.
The $k=2$ case belongs to an intermediate regime, which, within our accuracy,
we cannot classify as underdamped or overdamped.

Note that for $k=0,1$ the $G(k,t)/G(k,0)$ against $t/\beta^2$ curves
coincide (within our measurement accuracy) for all three values of $\beta$.
This coincidence, however, does not hold for higher values of $k$. It appears 
that the higher the temperature, the higher is $k$ at which the overdamping
sets in. In the next section we confront these properties of the gauge field
correlation functions with our rate measurements and with the ASY scenario.

There is no obvious way to compare our $G(0,t)$ with the $W$ real-time
correlation function computed in Ref. \cite{plas} for the Yang-Mills-Higgs 
theory. In that work the $k=0$ $W$ decay rate was found to be very small and
roughly proportional to $1/\beta$ in the high-temperature phase. However,
that result was obtained using local objects which only approximate gauge fields
(in the unitary gauge) sufficiently deep inside the low-temperature phase.

\section{Discussion}\label{disc}
We now summarize our findings. We saw that, with the spurious $N_{\rm CS}$
diffusion removed by the cooling method, the sphaleron rate in the 
low-temperature phase of the Yang-Mills-Higgs theory is too low to be measured
in the space-time volume attained in our simulations. The best we can do 
in this case is place an upper bound on the rate. That upper bound exceeds
the analytical sphaleron prediction by many orders of magnitude and hence it is
of limited value. The rate in the low-temperature phase close to the phase 
transition is a crucial parameter in any scenario of electroweak baryogenesis.
Our results, as well as those by Moore and Turok \cite{mt2}, indicate that a 
straightforward real-time simulation may not be an adequate tool for finding
this, very possibly extremely small, rate.

The focus of our numerical effort in this work is the approach of the classical
rate to a continuum limit in the Yang-Mills theory. Our results for large
volumes ($L>2\beta$) show that the dimensionless coefficient $\kappa$ tends
to decrease with the lattice spacing. This decrease, however, does not appear
to be rapid enough to match the ASY scenario, in which overdamping of gauge
fields at the magnetic length scale leads to the zero continuum limit for
$\kappa$. In this respect our results are in a qualitative agreement with
those of Moore and Turok \cite{mt2}, who used a different procedure for 
measuring 
topological charge. On the other hand, our values of $\kappa$ are systematically
lower than those of Moore and Turok, except for the point closest to the 
continuum limit, $\beta=16$.

We observe a much stronger dependence of $\kappa$ on the lattice spacing in
the small-volume setting $\beta=L$. Over a range of values of $\beta$, this
lattice spacing dependence is a simple proportionality. 

Our measurements leave open three possibilities for the behavior of the
transition rate close to the continuum limit.
\begin{enumerate}
\item The rate vanishes in the continuum limit, regardless of the lattice 
size to the magnetic screening length ratio. For small volumes this 
scenario is consistent with our observations. For large volumes, it can be
reconciled with our data if a more rapid decrease in $\kappa$ sets in at 
larger $\beta$.
\item Both the large-volume and the small-volume rates are finite in the 
continuum limit. This scenario is favored by our large-volume data. It could be
that an asymptotic approach of the small-volume $\kappa$ to the continuum limit
occurs at $\beta$ values larger than those considered here. 
\item The large-volume rate remains finite, but the small-volume rate vanishes
in the continuum limit\footnote{It seems very likely that $\kappa$ is a
growing function of the volume. Thus we discard here yet another possibility,
whereby the large-volume rate vanishes in the continuum, but the small-volume
one remains finite.}. This possibility is favored by our data.
\end{enumerate}
Consider now the additional information on the gauge field dynamics obtained
from the Coulomb-gauge correlation functions.
As we have seen, motion of gauge fields with momenta 0 and $2\pi/32$ is 
overdamped. A closer examination shows, however, that this is not quite the 
overdamping of long-wavelength modes predicted by ASY. Indeed the time scale
for the correlation function decay is proportional to $\beta^2$ for these
wavelengths, if $\beta$ is varied while keeping the wavelength {\it fixed}.
ASY predict that the decay time at a {\it fixed} wavelength should behave as
$1/\beta$. The decay times are short, of the order of several time units, 
{\it i.e.,} shorter than the corresponding wavelengths, whereas, in the ASY
derivation, it is essential that the decay times be much longer than the
wavelengths. Moreover, as our magnetic mass measurement shows, the overdamping
only occurs for modes with wavelengths somewhat above the magnetic screening
length. Harder modes, presumably corresponding to the magnetic scale, are not
overdamped.

We can now discuss the three possibilities we have listed for the rate
behavior in the continuum in the light of our knowledge of the gauge
field dynamics in the Coulomb gauge.
Consider sphaleron transitions in a small-volume Yang-Mills theory. With
periodic boundary conditions, the sphaleron solution extends over the entire 
volume of the system \cite{baaldas}. Therefore it has a significant 
zero-momentum component of the gauge field whose motion is overdamped, with
the decay time $\sim\beta^2$ rather than $\sim\beta=L$. It is possible that this
overdamping gives rise to the $1/\beta$ behavior of $\kappa$ in small volumes.
On the other hand, it appears unlikely that very soft 
overdamped modes play a role in the $N_{\rm CS}$ diffusion in large volumes. 
Indeed, for our $L=32$ lattices, these modes correspond to 
excitations whose linear size is close to $L$. If these modes were important
for formation of sphaleron-like configurations, we would expect a strong
finite-size dependence of the rate. Such a dependence, however, is not observed
({\it cf} the $\beta=14$ rate for $L=32$ and for $L=36$). Hence the 
sphaleron-like transitions, whose spatial size is determined by the magnetic
mass, should not be strongly affected by damping of these soft modes.
Modes with momenta on the magnetic scale or harder are not overdamped. Thus
it appears that the ASY mechanism does not operate in topology-changing 
processes in our large-volume simulations.

The argument just given supports the third possibility on our list. That 
possibilty, however, is quite unusual: it requires that $\kappa$ vanishes
above some critical value of $\beta/L$. A way to avoid this situation could
be as follows. Topology-changing transitions in small volumes could proceed
through barrier configurations whose size is considerably smaller than $L$.
These configurations need not have overdamped Fourier components. 
Indeed, all but the constant modes on our $\beta=L$ lattices are
not overdamped. These smaller barrier configurations
are stronger suppressed by Boltzmann factors than the 
sphaleron, but that does not necessarily preclude their finite contribution
to the rate in the continuum. That continuum, $\beta$-independent $\kappa$ 
might be so small that,
in the range of $\beta$ values considered, $\kappa$ is still dominated by the
$1/\beta$ term. This is a way in which the second possibility on our list
may be realized.

Finally, the first possibility on our list cannot be ruled out. While it is not
favored by our data, the case may be otherwise in a weaker-coupling regime.
In the current work the Debye to magnetic mass ratio is below 3.7, so these
two scales may not be completely separated. It may be necessary to further
increase $\beta$ in order to reach the range of validity of perturbation
theory on which the ASY scenario is substantially based. Since $m/m_D$ only
decreases as $1/\sqrt{\beta}$, while the rate decreases at least as rapidly as
$1/\beta^4$, exploring couplings much weaker than those presently considered
is likely to be computationally costly.

\section*{Acknowledgments}
We are indebted to L. McLerran  and R. Venugopalan for illuminating discussions.
Thanks are due to G.D. Moore, J. Smit, and M.E. Shaposhnikov for useful
comments. We also are grateful to Arnold, Son and Yaffe and to G.D. Moore
and N. Turok for communicating their results to us prior to publication;
to U.M. Heller, J. Rank, J. Hetrick and S. Gottlieb for advice and help with
Fourier-accelerated gauge fixing. Parallel implementation of algorithms used
in this work is based on software developed by the MILC collaboration
\cite{MILC_code}.
Numerical simulations reported in this work were performed on the SP/2 
massively parallel computer at the UNI-C.

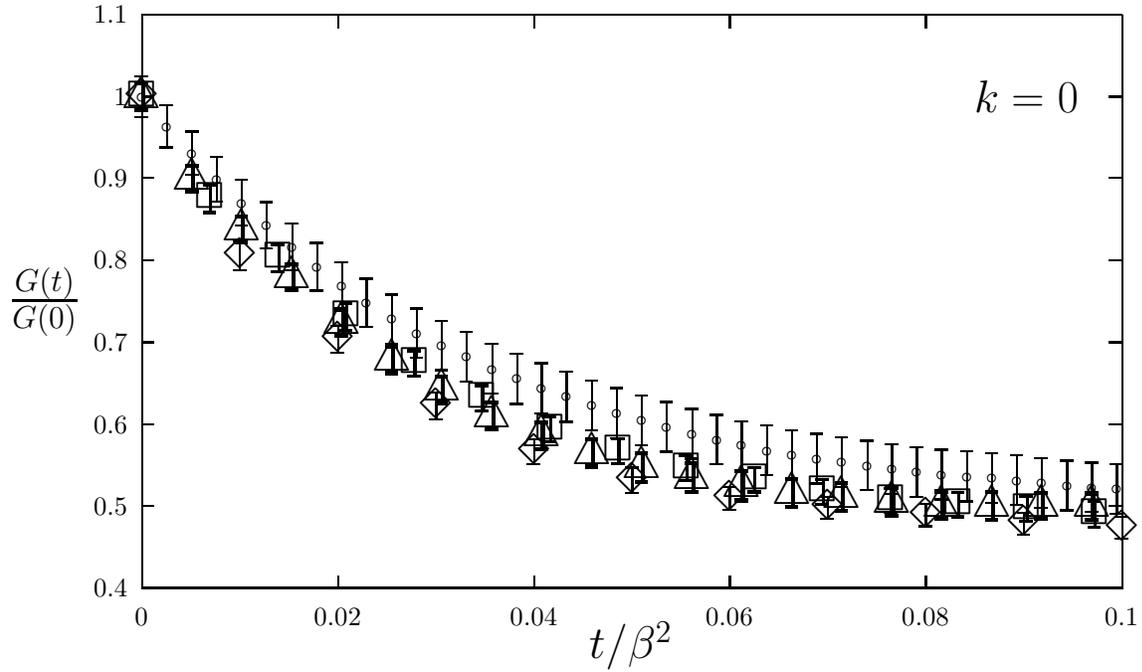
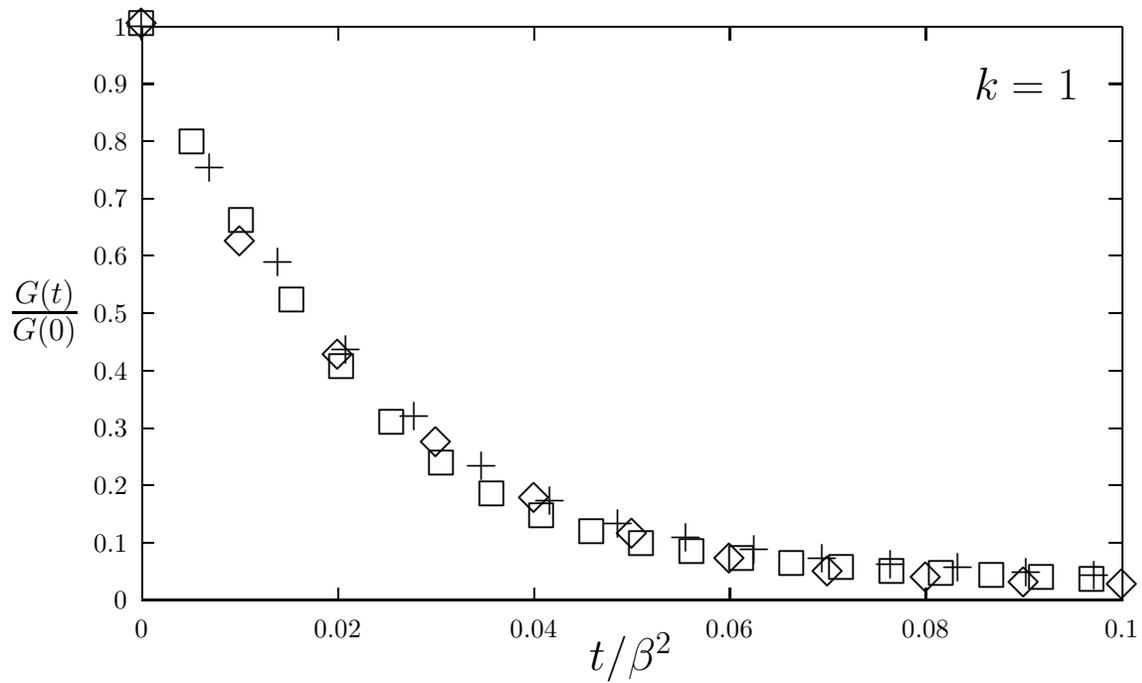
\end{document}